\documentclass[11pt]{article}

\usepackage[]{acl}

\usepackage{times}
\usepackage{latexsym}
\usepackage[T1]{fontenc}
\usepackage[utf8]{inputenc}
\usepackage{microtype}
\usepackage{inconsolata}
\usepackage{graphicx}
\usepackage{booktabs}
\usepackage{amsmath}
\usepackage{amssymb}
\usepackage{multirow}
\usepackage{xcolor}
\usepackage{tabularx}

\title{How Different, or Different How? Stance Granularity in Viewpoint Selection for Dialog Systems that Engage Users with Alternative Views}

\author{Anonymous ARR submission}

\author{%
  Morita Tarvirdians\thanks{\ Equal contribution.} \quad
  Reeve Lorena\footnotemark[1] \quad
  Catharine Oertel \\
  TU Delft, Delft, The Netherlands \\
  \texttt{M.Tarvirdians@tudelft.nl} \quad
  \texttt{reevelorena@gmail.com} \quad
  \texttt{c.r.m.m.oertel@tudelft.nl}
}
\begin{document}
\maketitle

\begin{abstract}
Systems that aim to help a user engage with viewpoints other than their own have to decide how different those viewpoints should be. Viewpoints are mainly represented on a scalar scale, too coarse to distinguish viewpoints that disagree with the user about different things. 
We introduce a selection policy in which a bandit sets how far viewpoints may sit from the user, and the policy selects the most novel viewpoint within that bound. We vary only its stance encoder, comparing a scalar score with Perspectivized Stance Vectors, which
break a position into separate stances on concepts. 
We integrate it inside a dialog system and run a between-subjects study (N~=~63). 
Concept-level stance produces more varied exposure sequences and broader coverage of the topic's positions, but does not change how deeply users justify viewpoints or how accurately they reconstruct them. 
Finally, we analyze user--viewpoint relations at the concept level, showing that users reconstruct a viewpoint least accurately on concepts that one side took no position on, rather than on concepts where the two disagreed.
Our findings open up opportunities for dialog systems that support engagement with alternative viewpoints by acting on which concepts a
user and a viewpoint disagree on, rather than on distance alone.

\end{abstract}

\section{Introduction}
\label{sec:introduction}

\begin{figure}[t]
  \centering
  \includegraphics[width=\columnwidth]{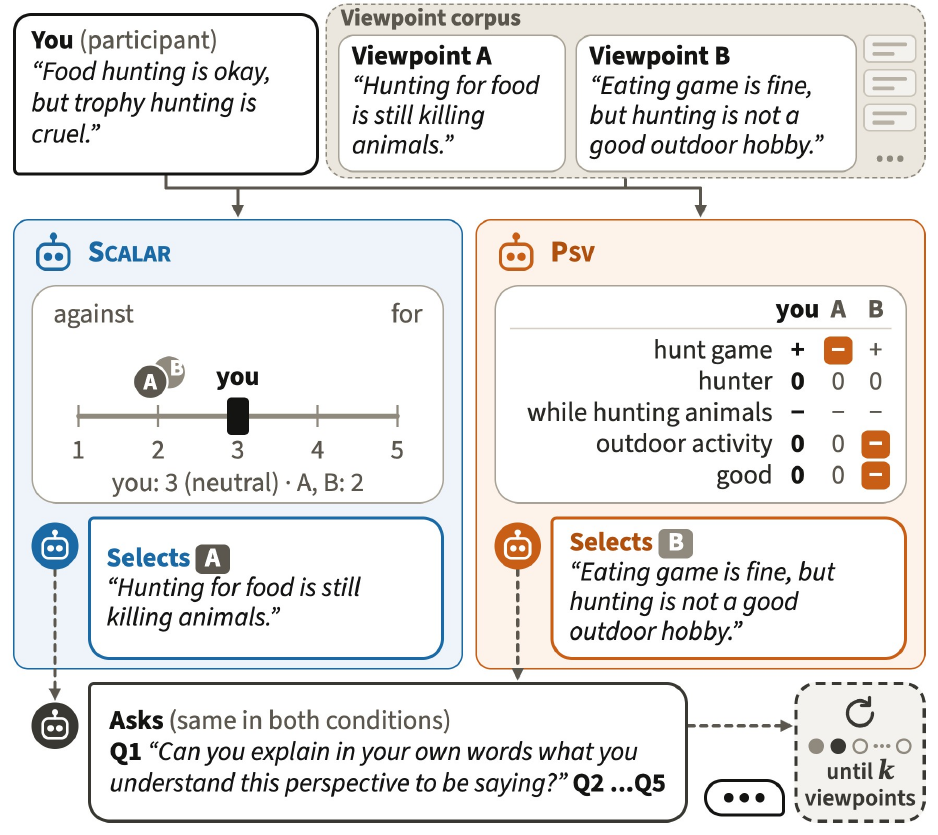}
  \caption{Two viewpoints a single stance score cannot tell apart. Under \textsc{Scalar} both are rated 2 against the user's 3, so neither is closer and the system picks arbitrarily. Under \textsc{Psv} they disagree with the user on different concepts (highlighted), so B is selected as the more novel.
  Both conditions then ask the same Socratic style questions, repeating until $k$ viewpoints are shown.}
  \label{fig:teaser}
\end{figure}

Public discourse increasingly takes place in information environments such as social media feeds, search results and recommender systems, where people selectively encounter views consistent with their existing positions \citep{frey_recent_1986,zarouali_overcoming_2021}. 
In politically contentious settings this is associated with affective polarization, which increases hostility toward opposing groups and reduces willingness to interact with them \citep{iyengar_fear_2015,iyengar_origins_2019}. 
One response is to structure exposure to perspectives other than a user's own, drawing on perspective-taking, the process of understanding a situation from another person's point of view \citep{reniers_editorial_2022}.

Providing alternative views is not sufficient on its own, because the relation between a new view and a person's existing position shapes how they process it \citep{lewandowsky_misinformation_2012,zhou_processing_2024}. 
A view close to what a user already holds adds little, while one far from it may be dismissed rather than considered \citep{sherif1961}.
Which views a user sees should therefore depend on where that user stands, and several systems select on this basis, choosing what to
show next from the arguments a user has already been exposed to \citep{aicher_towards_2022,aicher_self-imposed_2023,aicher_towards_2023}. 
Such a system compares each candidate viewpoint with the user's own position, and a viewpoint can differ from that position in two ways. It can differ in degree, by opposing the user more or less strongly, and in kind, by opposing a different aspect of the issue. Which of these differences a system selects on is an open question.

How a system represents stance of a viewpoint decides which of the two it can act on at all.
Stance is usually summarised as a single point on an ordinal scale running from opposing to supporting a position, which records degree and discards kind. 
Two viewpoints that oppose different aspects of an issue can land on the same point of that scale, and a system comparing them with a user sees one value for both.
Figure~\ref{fig:teaser} shows this for hunting. Both viewpoints sit two steps from the user's position, so a system using the \textsc{Scalar} scale has no basis for preferring either, even though the aspects they disagree with the user about are not the same. A system that cannot see a difference cannot select on it, whatever rule it applies.
 
Finer representations do exist. Perspectivized Stance Vectors (\textsc{Psv}), for example, record a separate stance toward each of several concepts in an issue \citep{plenz-etal-2025-argumentation}, so two viewpoints that oppose different aspects receive different vectors (Figure~\ref{fig:teaser}, right).
What such a representation changes for a user has not been tested, because a system that adopts a new representation normally changes the rule that uses it at the same time, leaving the two effects confounded. 
We therefore keep selection policy--the rule that decides which viewpoint to show next-- fixed and swap only the stance representation underneath it, comparing a single ordinal score (\textsc{Scalar}) with Perspectivized Stance Vectors (\textsc{Psv}). 
We implement this policy in a dialog system that presents the selected viewpoints and asks users about each of them.

Isolating the representation in this way lets us ask three questions. RQ1: \emph{How does representing stance at the concept level, rather than as a single score, change the viewpoints shown to a user?} RQ2: \emph{Does concept-level stance representation affect how fully users justify viewpoints and how accurately they reconstruct them?} RQ3: \emph{How does the concept-level relation between a user's stance and a viewpoint's stance relate to reconstruction accuracy?} 
Answering them contributes a viewpoint-selection policy that operates over interchangeable stance representations, an account of
what each representation makes available both to users and across the full stance space, and a concept-level analysis of reconstruction that a single aggregate score leaves hidden.
\section{Related Work}
\label{sec:related}

Selecting viewpoints for a particular user draws on three lines of work.
Argument retrieval defines how candidates are found.
User-conditioned selection makes that choice depend on the recipient and on what they have already seen.
Work on stance supplies the signal that relates a candidate to a user. 
We review each below, asking in each case how the relation between a user and a candidate viewpoint is represented, since that representation is what we vary.

\paragraph{Argument retrieval and user-conditioned selection.}
Argument retrieval finds arguments relevant to a query, including arguments with particular argumentative properties \citep{wachsmuth-etal-2018-retrieval}.
Argument recommendation conditions that selection on the recipient. Brenneis et al.~\citep{brenneis-etal-2021-will} introduce a dataset for evaluating argument recommenders from personal attitudes and interactions with arguments, more recent work incorporates user or stakeholder perspectives \citep{heinisch-etal-2024-tell, aicher_towards_2022}, and PerspectiveArg2024 formulates perspective-aware argument selection as a retrieval task \citep{falk-etal-2024-overview}. 
These approaches establish user and perspective information as retrieval signals, but differ in how they represent the relation between a user and a candidate argument.

Selection can also unfold over a conversation, with the user's current state and the arguments already shown affecting what comes next. 
Aicher et al.~\citep{aicher_towards_2022,aicher_towards_2023} model exposure histories and user positions in argumentative dialog systems to identify arguments that broaden subsequent exposure, showing that selection can operate as a policy over an evolving interaction rather than as one-shot retrieval. Our policy similarly selects over a session rather than in one shot.

\paragraph{Stance as a retrieval signal.}
Stance provides a direct signal for distinguishing arguments in controversial discussions. Methods exist for identifying the position expressed toward a topic or target \citep{durmus-etal-2019-determining,kobbe-etal-2020-unsupervised}, and
stance has been incorporated into retrieval and ranking through stance-aware query expansion \citep{kiesel-etal-2021-image} and stance-aware re-ranking for comparative retrieval \citep{reimer-etal-2023-stance}. 
These approaches treat stance as a property attached to an argument. In our setting stance instead defines the space in which candidates are compared, so that the distance between a user's stance and a candidate determines eligibility and the stance
representation determines which differences a policy can distinguish.

Conditioning on stance distance is supported by work on how arguments are processed. 
Social Judgment Theory models messages relative to a receiver's existing position \citep{sherif1961}, and studies of belief revision and misinformation show that the discrepancy between incoming information and prior beliefs affects how that information is processed \citep{durmus-cardie-2018-exploring,zhao2021,lewandowsky_misinformation_2012}.
This literature motivates conditioning selection on user--viewpoint distance without specifying how that distance should be represented.

\paragraph{Fine-grained stance representations.}
A compact stance representation characterizes an argument's position toward a target, but collapses distinctions between positions on different aspects of an issue. Plenz et al.~\citep{plenz-etal-2025-argumentation} address this with Perspectivized Stance Vectors, which represent an argument through its stances toward a shared set of issue-specific concepts, making agreement and
disagreement explicit at the level of individual concepts. They use concept-level stance to analyze agreement between arguments after retrieval. We use it instead as the representation over which a viewpoint-selection policy operates.

\section{Method}
\label{sec:method}

\begin{figure}[t]
    \centering
    \includegraphics[width=\columnwidth]{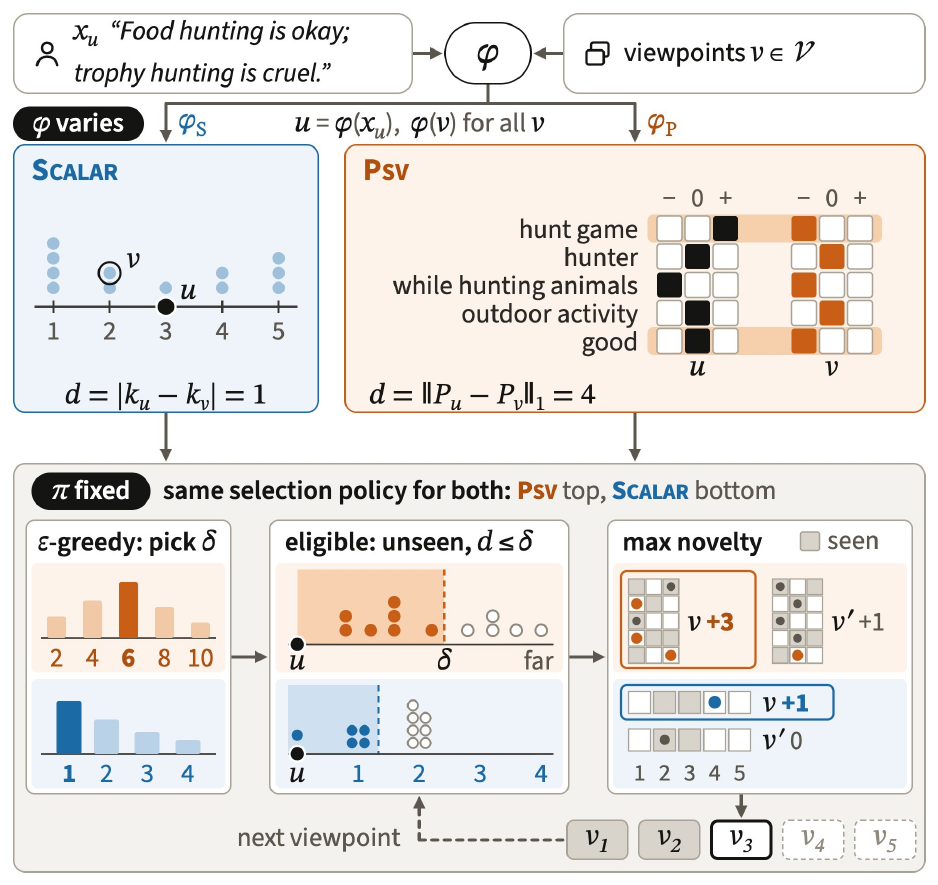}
    \caption{Viewpoint-selection framework. Only the stance encoder $\varphi$
    varies. It maps the participant stance and the candidate viewpoints into a
    scalar or a concept-level (PSV) space, giving the distance $d$. The policy $\pi$ is identical in both conditions. An $\varepsilon$-greedy mechanism sets the bound $\delta$, and the most novel unseen viewpoint within it is selected, repeating five times}
    \label{fig:policy}
\end{figure}

We formulate viewpoint selection as a retrieval problem over a fixed corpus of viewpoints. Our methodology represents the stance of the participant and candidate viewpoints, computes their distance in the resulting stance space, and uses this distance together with novelty to select and order five viewpoints. We integrate this selection policy into a dialog system that encodes the participant's free-text stance and presents the selected viewpoints. Figure~\ref{fig:policy} illustrates this selection framework. The remainder of this section describes the viewpoint corpus, stance representations, selection policy, and the dialog system.

\subsection{Viewpoint Corpus}
\label{sec:corpus}

We use the animal-hunting corpus of \citet{plenz-etal-2025-argumentation}, drawn from Debate.org. After  filtering, the corpus contains 364 viewpoints (Appendix~\ref{app:filtering}). The original corpus provides Perspectivized Stance Vectors (PSVs) over 100 concepts.
For selection we reduce this to a fixed five-concept signature, \emph{hunt game}, \emph{hunter}, \emph{while hunting animals}, \emph{outdoor activity} and \emph{good}. 
Concepts were ranked by how much their stance varies across the corpus, how often they take a non-neutral stance, and how well they separate pro from con arguments, and the top five were retained
(Appendix~\ref{app:signature}). 
The reduction preserves the structure of the full 100-concept representation, which yields a closely matching clustering of the corpus (adjusted Rand index $.92$).

\subsection{Stance Representations}
\label{sec:representations}

A \emph{stance representation} encodes a position so that two positions can be compared numerically. We compare a scalar representation with a concept-level representation based on PSVs.

\paragraph{\textsc{Scalar}.}
A stance is encoded as an ordinal value $k_a\in\{1,\ldots,5\}$, from strongly against to strongly for, following a common stance operationalization \citep{chuang2025}. The distance between two stances is
$d_{\textsc{s}}(a,b)=|k_a-k_b|$.
This representation distinguishes five overall positions but does not encode which aspects of the issue produce the difference.

\paragraph{\textsc{Psv}.}
Following \citet{plenz-etal-2025-argumentation}, a stance is encoded as an $N\times3$ one-hot matrix $P$, with each row representing a stance (against, neutral, or for) on concept $i$. Distance is the number of concept--stance changes between two positions:
\[
d_{\textsc{p}}(P_a,P_b)=
\sum_{i=1}^{N}\sum_{j=1}^{3}|P_{a,ij}-P_{b,ij}|.
\]
With $N=5$, the representation can encode $3^5=243$ positions, of which 35 occur in our corpus.

\subsection{Selecting and Ordering Viewpoints}
\label{sec:policy}

We formulate viewpoint selection as retrieval over a fixed corpus, because rather than generating new content, the system selects one unseen viewpoint at each step.
Let $\mathcal{V}$ be the corpus (Section~\ref{sec:corpus}), $\phi:\Sigma^*\rightarrow\mathcal{S}$ the stance encoder, and $d$ a distance over the resulting stance space (Section~\ref{sec:representations}). 
A participant’s stance is $u=\phi(x_u)$, where $x_u$ is their own view on the topic, written as free text before the session. 

Let $\mathcal{H}$ denote the viewpoints already shown. At each step, the policy first restricts selection to unseen viewpoints within a distance bound $\delta$ of the participant's stance. This defines the eligible set:
\begin{equation}
\mathcal{W} = \{v\in\mathcal{V}\setminus\mathcal{H} :
d(\phi(v),u)\leq\delta\}.
\label{eq:eligible}
\end{equation}

Distance alone does not determine which eligible viewpoint is selected. To favor viewpoints that add new stance information, we score each candidate by
the stance features it introduces that have not appeared in previously shown viewpoints:
\begin{equation}
\operatorname{nov}(v,\mathcal{H}) =
\Bigl|\sigma(v)\setminus\bigcup_{h\in\mathcal{H}}\sigma(h)\Bigr|,
\label{eq:novelty}
\end{equation}
where $\sigma(v)$ is the set of stance features associated with $v$. Under
\textsc{Psv}, these are the five concept--stance pairs; under \textsc{Scalar},
the feature is the single ordinal stance level.

The policy then selects the eligible viewpoint with the highest novelty score:
\begin{equation}
\pi(u,\mathcal{H}) = \operatorname*{arg\,max}_{v\in\mathcal{W}}
\operatorname{nov}(v,\mathcal{H}).
\label{eq:policy-argmax}
\end{equation}
Ties are broken by smallest distance and then uniformly at random. If no unseen viewpoint satisfies the distance bound, the nearest unseen viewpoint
is selected. Figure~\ref{fig:policy} illustrates these three stages.

Distance and novelty serve complementary purposes. The distance bound limits how far a viewpoint may differ from the participant, measured as the difference between overall ratings under \textsc{Scalar} and the $L_1$ distance between concept-level stance vectors under \textsc{Psv}. Novelty favors viewpoints that introduce stance features not covered by the session history. Without novelty, the policy could repeatedly select nearby viewpoints with highly similar positions.

The stance representation determines how both criteria operate.
\textsc{Scalar} encodes five overall positions and assigns one feature to each viewpoint, making novelty binary. At any given distance from the participant, at most two scalar positions are possible. \textsc{Psv} represents each viewpoint through five concept--stance pairs, allowing novelty to reflect which aspects of the issue have not yet been encountered. 

At the start of each session, an $\varepsilon$-greedy bandit \citep{sutton2018} selects $\delta$ from the distances available in that
condition. 
The bound remains fixed throughout all five viewpoints. The bandit's arm values are updated between sessions using the depth of participants' answers (Section~\ref{sec:measures}); responses do not change the sequence within the current session.

\paragraph{Dialog system.}
\label{sec:dialog}
The policy runs inside a text-based dialog system. Each turn presents a
viewpoint unchanged from the corpus, framed as the view of a person identified
by an arbitrary unisex first name. Five names are used, with no further
information provided about these individuals. The system then asks five fixed
Socratic-style questions \citep{danry2023,ang2023socratic} that scaffold
perspective-taking by asking the participant to articulate each position from
the speaker's point of view \citep{reniers_editorial_2022}. They move from
restating the position, through the reasons and assumptions behind it, to how it
relates to the participant's own view (Appendix~\ref{app:questions}).
Participants answer in free text. A session presents $k = 5$ viewpoints and
therefore 25 written answers from each participant. The bandit uses the depth
(Section~\ref{sec:measures}) of these answers to update its arm values,
informing the distance bound selected for subsequent sessions.
\section{User Study}
\label{sec:setup}

\subsection{Study Design and Procedure}
\label{sec:participants}
We conducted a between-subjects study on \emph{Should animal hunting be allowed?}.
The study was online.
Participants first completed a consent form, the perspective-taking subscale of the Interpersonal Reactivity Index (IRI-PT) \citep{davis1983}, and wrote their own view on the issue as free-text. Then they interacted with the dialog system (Section~\ref{sec:dialog}), followed by the reconstruction task (Section~\ref{sec:measures}).

\subsection{Participants}
Participants were auto-balanced between the two conditions. Of 79 participants, 63 completed the study, 31 in \textsc{Scalar} and 32 in \textsc{Psv} condition. They were recruited through personal networks and were not compensated. Conditions did not differ on IRI-PT, gender, or age bracket (Appendix~\ref{app:demographics}).

\subsection{Measures}
\label{sec:measures}

\paragraph{Exposure (RQ1).}
From session logs, we count distinct exposure sequences and the
concept--stance pairs covered per session. We also replay the policy over the
full corpus for every possible participant position.

\paragraph{Justification depth (RQ2).}
To measure whether participants gave reasons rather than merely restating a
viewpoint, each written answer is scored from 0 (bare assertion) to 3 (two or
more distinct reasons), following \citet{trenel2004,steenbergen2003}. This
measure reflects the role of addressing reasons in engaging with opposing
positions \citep{kuhn2007}. Scores are averaged across the five answers for
each viewpoint and rescaled to $[0,1]$.

\paragraph{Reconstruction accuracy (RQ2, RQ3).}
After each session, participants reconstruct one randomly selected viewpoint
by labeling each of its five concepts as against, neutral, or for. We compare
the reconstructed profile $\hat P$ with the true profile $P$ using
$1-L_1(\hat P,P)/(2N)$, yielding scores from 0 to 1, with chance at $1/3$.

\paragraph{Target specificity (RQ3).}
We score the reconstruction of one randomly selected viewpoint against all
five viewpoints shown in the session. Specificity is the score difference
between the target and the other viewpoints, adjusted by the score expected
from a reconstruction with no knowledge of the target. It therefore measures
whether the reconstruction identifies the intended viewpoint beyond what
could be explained by how distinct the session's viewpoints are.

\paragraph{Concept-level relation (RQ3).}
For each concept, we classify the participant's and viewpoint's stances as
matching, opposing, or involving at least one neutral stance. We also test
whether errors on differing concepts tend to reflect the participant's own
stance, comparing observed errors with a permutation null.

\subsection{Analysis}
\label{sec:analysisplan}

Only the condition comparison of reconstruction accuracy was pre-specified;
all other analyses are exploratory. We report estimates with confidence
intervals and apply Holm correction within test families. Per-concept outcomes
are analyzed using binomial GEE with participant-clustered standard errors,
and justification depth using a mixed model with participant random
intercepts. The corpus replay covers all participant positions and distance
bounds, with 100 seeds per setting.

\section{Results}
\label{sec:results}

\subsection{Concept-Level Stance Individualizes and Broadens Exposure}
\label{sec:rq1}

\begin{figure}[t]
\centering
\includegraphics[width=\columnwidth]{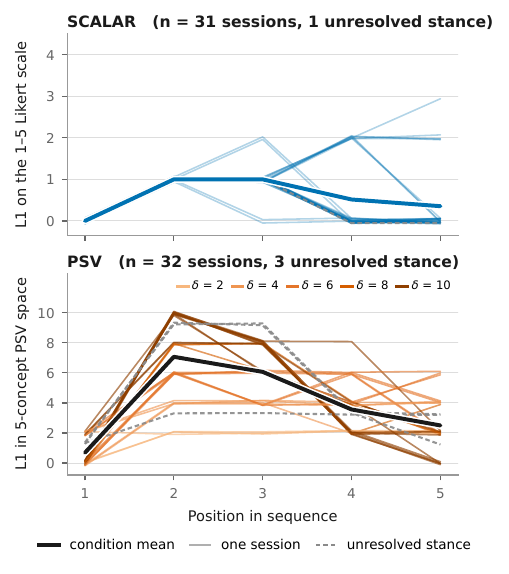}
\caption{Distance between each viewpoint and the participant by sequence position. Top: \textsc{Scalar}; bottom: \textsc{Psv}. Scales differ and should not be compared vertically. Lines show sessions; thick lines are condition means, dashed lines the four sessions with unresolved participant stance, and jitter separates identical sequences.}
\label{fig:trajectories}
\end{figure}

RQ1 asks how concept-level stance representation changes the viewpoints shown to participants.
Figure~\ref{fig:trajectories} shows each viewpoint's distance from the participant in the representation used for selection. 
Because the panels use different scales, we compare counts rather than distances across conditions. We count distinct distance sequences, each defined by the five distances in presentation order; sessions with the same distance sequence count as one even if they show different viewpoints.

\paragraph{Sequences.}
The 31 \textsc{Scalar} sessions produced 6 distinct distance sequences, compared with 24 across the 32 \textsc{Psv} sessions. Under \textsc{Scalar}, the first viewpoint matched the participant's stance in all 31 sessions, the second was one step away in all 31, and the third was one step away in 27. The fourth and fifth returned to the participant's stance in 23 and 26 sessions, respectively.
The sequence is fully determined by the participant's stance and the distance bound, with each of the 9 observed combinations
producing exactly one sequence (Appendix~\ref{app:rq1}). Over the condition as a whole, 82 of the 155 viewpoints shown (52.9\%) sat at the participant's own stance.
Under \textsc{Psv} the first viewpoint matched the participant on all five concepts in 20/32 sessions. The
second sat at the furthest distance the bound allowed in 27/32, after which distance fell across the remaining slots, with means of 7.1, 6.1, 3.6 and 2.5 at slots 2 to 5.

\paragraph{Coverage.}
The issue has 15 concept--stance combinations, one for each of the three stances on each of the five concepts, and a session covers those its viewpoints take between them. 
\textsc{Psv} sessions covered 13.7 of the 15 on average (SD 2.2), against 11.0 under \textsc{Scalar} (SD 2.4). All 15 appeared in 22 of 32 \textsc{Psv} sessions and in 1 of 31 under \textsc{Scalar}.

\paragraph{Distance and coverage.}
At each sequence position, participant distance was strongly correlated with the number of newly introduced concept--stance combinations under both \textsc{Psv} ($\rho=.72$--$.95$) and \textsc{Scalar} ($\rho=.73$--$1.00$), where both measures varied. 
Thus, more distant viewpoints tended to introduce more new combinations, linking exposure to disagreement with the novelty criterion used by the policy.

\paragraph{Corpus replay.}
To test whether these patterns depended on the participants sampled, we replayed the deployed policy across the full corpus for all possible participant positions, 5 scalar ratings and 243 \textsc{Psv} profiles. We evaluated every distance bound using 100 seeds. Each \textsc{Scalar} rating--bound combination produced exactly one sequence. 
At the widest bound, \textsc{Psv} covered all 15 combinations in every run, compared with a mean of 12.7 (SD 1.2) under \textsc{Scalar}, which covered all 15 in only 5\% of runs (Appendix~\ref{app:results}). At the narrowest bound, mean coverage was 9.1 under \textsc{Psv} and 8.3 under \textsc{Scalar}; \textsc{Psv} selected the nearest unseen viewpoint as a fallback in 52.5\% of selections. The corpus contained at most two scalar rating levels at any given distance, compared with a median of 6--12 \textsc{Psv} profiles at each $L_1$ distance from 4 to 8.

Overall, concept-level stance representation produced more varied distance sequences and broader coverage of concept--stance combinations. The distance bound constrained how far selected viewpoints could differ from the participant, while novelty shaped which viewpoints were selected within that constraint.

\subsection{Reconstruction Is Hard Under Both Representations}
\label{sec:rq2}

RQ2 asks whether concept-level stance affects justification depth and viewpoint reconstruction.

Participants reconstructed 43.9\% of concept stances under \textsc{Scalar} (68/155) and 48.7\% under \textsc{Psv} (78/160), OR 1.22 [0.72, 2.06]. Both conditions sit above the 33.3\% chance level. 
We re-ran the comparison without the 21 ``not sure'' answers, without the four participants whose stance was unresolved, without each participant in turn, and against the second labelling of the same viewpoints. The odds ratio stayed between 1.10 and 1.41 throughout, so the small \textsc{Psv} advantage is steady and never becomes reliable (Appendix~\ref{app:agreement}).

We score each participant's description not only against the viewpoint they were asked about but also against the other four they saw. It fitted the intended viewpoint best for 24 of 32 participants under \textsc{Psv} and 13 of 31 under \textsc{Scalar}. That gap opens more easily when a session's five viewpoints are far apart, because one description can then fit only one of them, while in a session of similar viewpoints the same description fits all five about equally well. 
\textsc{Scalar} sessions were the tighter ones (mean distance 5.3 against
6.3). 
The size of the advantage does not differ between conditions (+0.055 [$-0.07$, 0.18]), and subtracting what a reconstructor with no knowledge of the target would score on the same sessions leaves it there
(Appendix~\ref{app:geometry}).

Justification depth, scored on the answers participants wrote during the session, showed no condition difference either. It averaged 0.44 under \textsc{Scalar} and 0.40 under \textsc{Psv} ($b=-.04$ [$-.15$, $.07$], 1,575 turns), with similar response length (30.7 against 28.6 words) and no condition difference on any question, although the common-ground question (Q5, Appendix~\ref{app:questions}) was the hardest in both conditions (0.19 against 0.60 on the question asking for reasons).

Across all 63 participants, the longer someone spent answering during the session, the better their reconstruction accuracy afterwards (OR 1.45 per SD of time on task [1.09, 1.93]). 
However, justification depth during the session fell from one viewpoint to the next, by 0.013 of the scale per slot, or about 0.05 between the first viewpoint and the fifth ($p=.001$). It did not depend on how far a viewpoint sat from the participant's own stance in either condition, so participants did not write shallower answers about the viewpoints they disagreed with most.

Stance representation therefore changed which viewpoints participants saw (Section~\ref{sec:rq1}) without changing how well they understood them.

\subsection{Neutral Positions, Not Disagreement, Are Hardest to Reconstruct}
\label{sec:rq3}

Participants were right on 55.6\% of concepts where they and the viewpoint took
the same side (79/142), 51.2\% where they took opposite sides (22/43), and
34.9\% where at least one side was neutral (44/126), against chance at 33.3\%
(Table~\ref{tab:relation}). Accuracy was higher where they agreed than where
either side was neutral (OR 2.09 [1.20, 3.65], Holm $p=.029$), and no different
where they disagreed (OR 1.19 [0.59, 2.42], $p=.63$). Disagreeing with a
viewpoint therefore did not make it harder to read. 

The shortfall sat where the participant was neutral, at 24\% (11/45), below
chance and below agreement (OR 0.28 [0.13, 0.61], $p=.001$). Where only the
viewpoint was neutral it was 42\% (22/53), no different from agreement.
Participant-neutral labels were also the least stable when re-measured, 10 of 25
changing, so we treat this split as exploratory. The pattern held in both
conditions ($p=.85$) and under offline re-measurement (OR 1.86, Holm
$p=.048$).

\begin{figure}[t]
\centering
\includegraphics[width=\columnwidth]{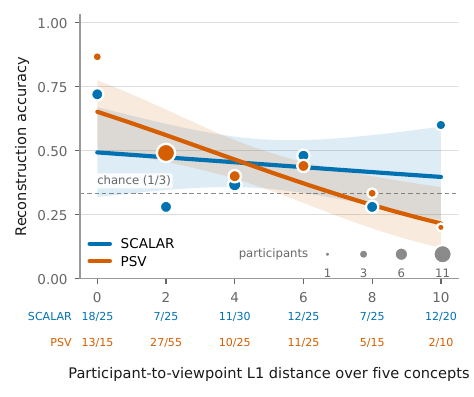}
\caption{Reconstruction accuracy by participant--viewpoint $L_1$ distance over
five concepts. Marker area indicates the number of participants at each
distance, lines are binomial GEE fits with 95\% bands, and the dashed line marks
chance (1/3).}
\label{fig:distance}
\end{figure}

Within \textsc{Psv}, accuracy fell as participant and viewpoint differed on more
concepts, OR 0.82 per $L_1$ unit [0.72, 0.92], about 0.67 per concept, over 29
participants (Figure~\ref{fig:distance}). The slope stayed between 0.74 and 0.88
across adjustments for text features, participant extremity, slot, offline
stance labels, concept relation, and dropping each participant in turn. It was
steepest where participant and viewpoint agreed (OR 0.68 [0.50, 0.93] against
0.98 where either was neutral), so distance cost accuracy even where they
agreed. Under \textsc{Scalar} we found no such association (OR 0.96
[0.84, 1.11]). The interaction was not reliable ($p=.33$) and the two distance
scales are not comparable, so this is not a condition difference.

\begin{table}[t]
\centering
\small
\setlength{\tabcolsep}{4pt}
\begin{tabular}{lrrr}
\toprule
\textbf{Relation} & \textbf{Scalar} & \textbf{Psv} & \textbf{All} \\
\midrule
Same side & 35/65 (.54) & 44/77 (.57) & 79/142 (.56) \\
Opposite sides & 9/19 (.47) & 13/24 (.54) & 22/43 (.51) \\
Either neutral & 24/70 (.34) & 20/56 (.36) & 44/126 (.35) \\
\quad participant & 5/28 (.18) & 6/17 (.35) & 11/45 (.24) \\
\quad viewpoint & 11/27 (.41) & 11/26 (.42) & 22/53 (.42) \\
\quad both & 8/15 (.53) & 3/13 (.23) & 11/28 (.39) \\
\bottomrule
\end{tabular}
\caption{Reconstruction accuracy by participant--viewpoint stance relation,
shown as correct/total (accuracy). Chance is .33. Participant stances were
extracted at runtime under \textsc{Psv} and post hoc under \textsc{Scalar}.}
\label{tab:relation}
\end{table}

Where participant and viewpoint agreed, 33 answers were still wrong. Of those,
21 said the viewpoint held no position and 12 gave the opposite one. Nine of the
21 were ``not sure'' answers, and without them the split is even, so we cannot
say participants tended to leave a stance unattributed rather than reverse it.

Participants might have substituted their own stance for the viewpoint's when reconstructing it. When their stance differed from the viewpoint's, two incorrect labels were possible, only one of which matched their own stance, giving a chance rate of 0.5. Participants used their own stance in 21 of 35 \textsc{Psv} errors and 23 of 48 \textsc{Scalar} errors. These proportions were close to the permutation-null rates of 47\% in both conditions ($p=.063$ and $p=.439$, respectively). The \textsc{Psv} result points in the predicted direction, but the difference amounts to only about four errors across 19 participants. With so few errors, we cannot distinguish this pattern from chance, so we treat it as inconclusive.

\section{Discussion}
\label{sec:discussion}

We investigated whether the stance representation a policy uses to select viewpoints changes what a user encounters and what they take away from it.

In \textbf{RQ1}, we investigated whether representing stance at the concept level, rather than as a single score, changes which viewpoints are selected. It does.
Both conditions ran the same selection policy, so the difference came from the encoder alone, \textsc{Psv} produced 24 distinct exposure sequences against 6 under \textsc{Scalar}, and participants encountered more of the concept--stance combinations available on the topic. Replaying the deployed policy over the full corpus and all possible participant positions shows the same pattern.
While prior work has either surfaced diverse opinions without selecting relative to the user \citep{Freelon01072012,kim2021starry}, or selected relative to the user while treating the stance encoding as given \citep{aicher_towards_2022,aicher_towards_2023}, our results indicate that the stance representation itself shapes selection behavior by defining which distinctions the policy can make between viewpoints.

Our second research question (\textbf{RQ2}) asked whether this difference in exposure carried through to how deeply participants justified the viewpoints they encountered and how accurately they reconstructed them. It did not, on either measure, but our study was powered to detect only large effects. 
This is consistent with \citep{saveski2022}, who found that exposure to opposing political content did not improve participants' reported understanding of why others held those views. 
In our study, exposure was deliberately structured and every viewpoint was followed by fixed Socratic style questions, yet reconstruction still did not differ between conditions. 
Showing participants a wider range of viewpoints did not help them reconstruct those viewpoints more accurately.

In \textbf{RQ3}, we investigated which parts of a viewpoint participants reconstructed correctly, and how this depended on their own stance. Participants were about as accurate on concepts where they opposed the viewpoint as on concepts where they agreed with it, and markedly less accurate where either side held no position.
Accuracy declined as the overall stance distance increased, consistent with the discrepancy account of \citet{sherif1961}. However, the most frequent reconstruction error was to label an opposing stance as neutral, and neutral positions were the hardest to reconstruct. This suggests that stance discrepancy may affect not only how a viewpoint is judged, but also how its specific claims are understood and remembered.

\paragraph{Implications}
These findings have implications for dialog systems that select viewpoints for a user. Breadth and stance discrepancy should be treated separately, since novelty relative to the user can favor the most distant viewpoints. A policy could instead target a particular stance distance and select the most novel viewpoint within that range, provided the representation distinguishes multiple viewpoints at the same distance. Our policy keeps the distance bound fixed within each session. Future work could explore within-session adaptation, updating this bound in response to user interaction to balance novelty and discrepancy dynamically.

\paragraph{Societal Impact}
\label{sec:ethical}
Polarization on contested issues is a long-standing societal problem. With this work we aim to contribute to supporting users in encountering and engaging with viewpoints other than their own.
We integrate this support into a dialog system, a type of system many people are now familiar with, and show that the stance representation a policy uses determines which disagreements a user is able to meet. 
This opens up opportunities for dialog systems that broaden what users encounter on contested issues while remaining accountable for which
dimensions of disagreement they act upon on.
\section{Conclusion}
\label{sec:conclusion}
We presented a viewpoint-selection policy that operates over interchangeable stance representations, allowing us to vary the representation while keeping the selection rule fixed. Evaluating it with 63 participants and by replaying it across the full stance space, we showed that representation shapes which differences the system can act on. Concept-level selection produced more varied exposure and broader topic coverage, but did not change justification depth or overall reconstruction accuracy. Concept-level analysis showed that participants reconstructed neutral concepts less accurately than concepts on which they agreed or disagreed with a viewpoint, a distinction obscured by overall accuracy.
Our findings establish stance representation as a design choice in viewpoint selection. This opens a direction for systems that control how far a viewpoint is from a user separately from which specific disagreements it raises.

\section*{Limitations}
\label{sec:limitations}
Our findings have several limitations. First, the study focused on a single contested issue, one language, and a corpus of user-contributed viewpoints, so the results may not generalize to other topics or content sources. 
Second, the concept-level representation used a fixed signature of five concepts selected during system development; different concepts or issues may produce different selection patterns. 
Third, reconstruction required participants to report a viewpoint's stance on each concept, which may conflate their understanding of its position with their recall of its content. 
Finally, the study involved 63 participants and was powered to detect only large effects in the condition comparison. The absence of differences in justification depth and reconstruction accuracy therefore does not establish that the representations are equivalent on these outcomes.

\section*{Ethics Statement}
\label{sec:ethics}

This study was approved by the Human Research Ethics Committee of our university. All participants provided consent before participation.
The dialog system offered no opinions or advice of its own and generated no arguments. 
That said, both selecting viewpoints close to the user's position and selecting ones far from them can have their own consequences, and deployment of our proposed approach on more controversial and sensitive topics and in real-world settings may therefore need further ethical considerations and approvals.

\section*{Gen AI Declaration}
Generative AI tools were used to assist with copy-editing and coding, but the authors are fully accountable for the content and ideas.

\section*{Acknowledgments}
We gratefully acknowledge the support of the Delft AI Initiative and the Designing Intelligence (DI) lab.
This research was partially funded by the Hybrid Intelligence Center, a 10-year programme funded by the Dutch Ministry of Education, Culture and Science through the Netherlands Organisation for Scientific Research, \href{https://www.hybrid-intelligence-centre.nl}{Hybrid
Intelligence Centre}, grant number 024.004.022.

\bibliography{custom}

\appendix
\newpage
\section{Corpus Quality Filtering}
\label{app:filtering}

\begin{table*}[t]
  \centering
  \small
  \begin{tabular}{llcl}
    \toprule
    & Description & Flagged & Pattern \\
    \midrule
    A & Incoherent or gibberish & 5  & \verb!(\b\w\b[ \t]+){8,}! \\
    B & External URL references & 5  & \verb!https?://|www\.! \\
    C & Strong profanity        & 7  & \verb!\bfuck\w*\b|\bshit\w*\b|\basshole\w*\b! \\
    D & Mild profanity          & 17 & \verb!\bdamn\b|\bhell\b|\bcrap\b|\bbitch\w*\b|\bbastard\w*\b! \\
    E & Repetitive meme or spam & 5  & \verb!(\b\w+[ \t]+\w+[ \t]+\w+[ \t]+\w+[ \t]+\w+\b)(?:.*?\1){3,}! \\
    F & Student essay artefacts & 3  & \verb!Name:.*?Teacher:|Hunting Essay|Class Period:! \\
    G & Excessive whitespace    & 1  & \verb!\n{5,}| {15,}! \\
    H & Repeated symbols        & 2  & \verb!(>[>\s]*){6,}|\*{4,}! \\
    \bottomrule
  \end{tabular}
  \caption{Corpus quality-filtering criteria, the number of entries each flagged, and the regular expression used.}
  \label{tab:filtering}
\end{table*}

Eight deterministic criteria were applied to the 408 raw arguments.
Table~\ref{tab:filtering} lists them with the number of entries each flagged and the pattern used.
Criteria B--F are case-insensitive, and E and F also match across line breaks. The criteria are not mutually exclusive, so the
per-criterion counts sum to 45 while 41 unique entries were flagged. 
Three further entries were gibberish that no criterion matched, because criterion A requires eight or more consecutive single-character tokens and the noise in these entries is interleaved with longer ones. We removed these three by hand and record their content hashes in the filtering script, so the reduction from 408 to 364 is reproducible.

\newpage
\section{Choosing the Five Concepts}
\label{app:signature}

Concepts were ranked by three criteria. 
\emph{Stance variance} ranks concepts by the variance of their discrete stance values~\citep{guyon2003}. 
\emph{Non-neutrality rate} ranks by the proportion of arguments taking a non-neutral stance, which operationalises the granularity principle of \citet{plenz-etal-2025-argumentation}. 
The
\emph{Fisher discriminant} quantifies each concept's power to separate the pro
and con groups, $F_i = (\mu_{\text{PRO},i} - \mu_{\text{CON},i})^2 / (\sigma^2_{\text{PRO},i} + \sigma^2_{\text{CON},i} + \varepsilon)$ \citep{hart2001}.
All five deployed concepts rank within the top nine by stance variance and the top ten by the Fisher criterion.

We evaluate the deployed signature against a reference clustering in the full 100-concept space. Arguments were clustered with hierarchical average linkage under $L_1$ distance, and the number of clusters was chosen by maximising the silhouette score \citep{rousseeuw1987} over $K \in \{2,\dots,6\}$, which selects $K = 2$. Agreement between the reduced-space and full-space partitions, measured by the Adjusted Rand Index \citep{hubert1985}, is 0.92 on the filtered corpus and 0.88 on the raw corpus.

Selecting the top concepts by any single criterion performs worse at the same size. 
Fisher-based selection at $N = 5$ reaches ARI 0.84 on the raw corpus, against 0.88 for the deployed signature, and stance variance ranking needs six concepts to reach 0.90.

\newpage
\section{Question Bank}
\label{app:questions}

Table~\ref{tab:questions} gives the five question templates verbatim. 
The order is fixed and identical in both conditions.

\begin{table}[h]
  \centering
  \small
  \begin{tabularx}{\columnwidth}{lX}
    \toprule
    & Template \\
    \midrule
    Q1 & Can you explain in your own words what you understand this
         perspective to be saying? \\
    Q2 & Why do you think someone might hold this view about animal
         hunting? \\
    Q3 & What evidence or assumptions might underlie this perspective? \\
    Q4 & How does this perspective compare to your own view on
         \{topic\}? \\
    Q5 & What common ground, if any, do you see between this perspective
         and your own? \\
    \bottomrule
  \end{tabularx}
  \caption{The five question templates, in presentation order.}
  \label{tab:questions}
\end{table}

Q2 to Q5 are shown alone, with only the topic substituted. Q1 is shown as part of a longer (first) message that introduces the viewpoint and then poses Q1.

Viewpoints are named Alex, Jordan, Morgan, Riley and Casey in order of appearance. 

\newpage
\section{Prompts}
\label{app:prompts}

Figure~\ref{fig:prompts} reproduces the three runtime prompts as they appear
in the deployed code. All are zero-shot, sent as a single user message with no
system prompt and no exemplars.
All three use \texttt{claude-haiku-4-5-20251001} with temperature unset, so at the API default. 

\begin{figure*}[t]
{\small\begin{verbatim}
Concept stance classifier (Psv), one call per concept:

Given a controversial topic discussed by a given argument, and an aspect, provide an
one-word answer to the following question: Considering a person writing this argument,
what is their attitude towards the given aspect: negative, neutral, or positive?

Topic: {topic}
Argument: {argument_text}
Aspect: {aspect}


Scalar stance rating, one call per participant and once per corpus argument offline:

Rate the overall stance of the following text towards the topic on a scale of 1-5:
  1 = Strongly against
  2 = Against
  3 = Neutral
  4 = For
  5 = Strongly for

Topic: {topic}
Text: {stance}

Reply with only the single digit (1, 2, 3, 4, or 5).


Justification-level scorer, one call per participant turn in both conditions:

You are scoring a participant's response using Trenel's (2004) justification level scale.

The participant was asked: {question_text}
They responded: {clean_response}

Justification level scale:
0 - No justification: simple assertion with no supporting reasoning
1 - Inferior justification: reasons given but not logically linked to the claim
2 - Qualified justification: explicit logical link between reason and claim
3 - Sophisticated justification: two or more distinct reasons for a single claim

Reply with only the integer 0, 1, 2, or 3.
\end{verbatim}}
\caption{The three runtime prompts.}
\label{fig:prompts}
\end{figure*}


\newpage
\section{Participant Demographics}
\label{app:demographics}

Age, gender, and the perspective-taking subscale of the Interpersonal Reactivity Index (IRI-PT)
\citep{davis1983} were collected in the pre-task survey.
Table~\ref{tab:demographics} reports all three for the 63 analysed participants.

Gender and age bracket are tested with the Fisher--Freeman--Halton exact test, which extends Fisher's exact test to $r \times 2$ tables, and IRI-PT with a Welch two-sample $t$-test. 
With these cell counts the tests have limited power.

\begin{table}[t]
\centering
\small
\begin{tabular}{@{}lrr@{}}
\toprule
 & \textsc{Scalar} & \textsc{Psv} \\
 & ($n=31$) & ($n=32$) \\
\midrule
\textit{Gender} ($p=.79$) & & \\
\quad Woman & 11 (35.5) & 10 (31.3) \\
\quad Man & 18 (58.1) & 20 (62.5) \\
\quad Non-binary & 1 (3.2) & 0 (0.0) \\
\quad Not reported & 1 (3.2) & 2 (6.3) \\
\addlinespace
\textit{Age bracket} ($p=.34$) & & \\
\quad 19--24 & 13 (41.9) & 14 (43.8) \\
\quad 25--34 & 13 (41.9) & 13 (40.6) \\
\quad 35--44 & 4 (12.9) & 1 (3.1) \\
\quad 45--54 & 1 (3.2) & 4 (12.5) \\
\addlinespace
\textit{IRI-PT}, 0--28 ($p=.65$) & & \\
\quad Mean (SD) & 15.5 (4.2) & 16.0 (4.1) \\
\quad Range & 8--25 & 9--24 \\
\bottomrule
\end{tabular}
\caption{Condition balance for the 63 analysed participants, as counts with
column percentages. No participant selected a bracket above 45--54. $p$ values
are Fisher--Freeman--Halton exact tests for gender and age, with declined
responses (not reported) excluded, and a Welch $t$-test for IRI-PT ($t(60.9)=0.46$).}
\label{tab:demographics}
\end{table}

\newpage
\section{Determinism of the Scalar Sequences}
\label{app:rq1}

Section~\ref{sec:rq1} reports that under \textsc{Scalar} the sequence of distances follows from the participant's stance and the distance bound. This appendix derives that sequence and checks it against the sessions.

Let $s$ be the participant's stance level and $\delta$ the bound. Under \textsc{Scalar} a viewpoint carries one stance feature, its level, so novelty is binary and the policy of Section~\ref{sec:policy} reduces to three steps. First, the eligible levels are those within $\delta$ of $s$. Second, the policy shows an eligible level that has not been shown yet, taking the nearest one when several remain. Third, once every eligible level has been shown, novelty is zero for all candidates and the tie-break on distance returns the level nearest $s$,
which is $s$ itself. Several corpus viewpoints share each level, so which text a slot shows varies, but its distance does not.

Table~\ref{tab:scalar-rule} lists every combination of $s$ and $\delta$ in the data with the observed sequence and the one this derivation gives. The two agree in all 9 combinations, covering 31 of 31 sessions and 155 of 155 slots.

\begin{table}[h]
  \centering
  \small
  \begin{tabular}{ccccc}
    \toprule
    $s$ & $\delta$ & $n$ & observed & derived \\
    \midrule
    1 & 1 & 2  & 0,1,0,0,0 & 0,1,0,0,0 \\
    2 & 1 & 4  & 0,1,1,0,0 & 0,1,1,0,0 \\
    3 & 1 & 10 & 0,1,1,0,0 & 0,1,1,0,0 \\
    4 & 1 & 5  & 0,1,1,0,0 & 0,1,1,0,0 \\
    1 & 2 & 2  & 0,1,2,0,0 & 0,1,2,0,0 \\
    2 & 2 & 3  & 0,1,1,2,0 & 0,1,1,2,0 \\
    3 & 2 & 3  & 0,1,1,2,2 & 0,1,1,2,2 \\
    3 & 3 & 1  & 0,1,1,2,2 & 0,1,1,2,2 \\
    4 & 4 & 1  & 0,1,1,2,3 & 0,1,1,2,3 \\
    \bottomrule
  \end{tabular}
  \caption{Observed and derived distance sequences under \textsc{Scalar}, by
  participant stance $s$ and distance bound $\delta$. $n$ is the number of
  sessions in each combination.}
  \label{tab:scalar-rule}
\end{table}

Two things limit what the agreement shows. First, every combination holds a single observed sequence, so the evidence is 9 combinations rather than 31 sessions, and two of them hold one session each. Second, slots 1 and 2 are identical in every session and therefore carry no information about the derivation.

\newpage
\section{Simulation Results}
\label{app:results}

The exposure results in Section~\ref{sec:rq1} hold for the stances and bounds the 63 participants happened to produce. To separate the policy from that sample, we replay the deployed selection code over the full corpus for every participant stance each representation admits, the 5 scalar positions and the 243 \textsc{Psv} profiles, every distance bound available in that condition and 100 random seeds per setting. Each run returns the five viewpoints the policy would select. We score a run by coverage, the number of the 15 concept--stance
combinations its five viewpoints take between them, which is counted the same way in both conditions and is therefore comparable across them.

Coverage rises with the bound in both conditions. Under \textsc{Psv} it reaches all 15 combinations before the bound is exhausted. Under \textsc{Scalar} it flattens below 15, because the corpus holds at most two scalar positions at each distance from the participant, so a wider bound adds few further positions to choose from.

\begin{figure}[h]
  \centering
  \IfFileExists{figures/fig2_coverage_by_bound.pdf}{%
    \includegraphics[width=\columnwidth]{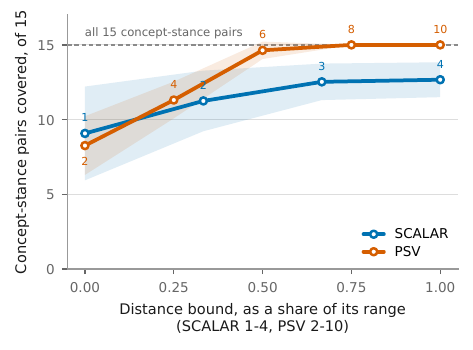}}{%
    \fbox{\parbox{0.9\columnwidth}{\centering figures/fig2\_coverage\_by\_bound.pdf
    missing}}}
  \caption{Simulated coverage of the 15 concept--stance combinations by
  distance bound, over the full corpus and every possible participant
  position, 100 seeds per setting. Each condition's bound is rescaled to its
  own range (\textsc{Scalar} 1--4, \textsc{Psv} 2--10) and labelled with its
  actual value. Bands show $\pm 1$ SD across runs. The two horizontal scales
  are not commensurable, so only the shapes of the lines should be compared.}
  \label{fig:coverage}
\end{figure}

\section{Label Agreement}
\label{app:agreement}

Participant stance profiles and corpus labels are both assigned automatically (Section~\ref{sec:representations}), so we check them in two ways. First, stability. We run the extraction twice over the same participant texts and compare the two profiles, 157 concept cells over the 32 \textsc{Psv}
participants, with three cells excluded where one run returned no decision. Second, agreement with a second labelling. 
We label the 47 viewpoints that served as reconstruction targets a second time and compare the result to the
corpus labels, 235 concept cells. Table~\ref{tab:label-agreement} reports Cohen's $\kappa$ per concept for both checks.

\begin{table}[h]
  \centering
  \small
  \begin{tabular}{lccc}
    \toprule
    Concept & Participant & Viewpoint & Neutral \\
    \midrule
    hunt game             & 0.84 & 0.74 & 6 \\
    hunter                & 0.81 & 0.78 & 7 \\
    while hunting animals & 0.66 & 0.74 & 6 \\
    outdoor activity      & 0.72 & 0.52 & 12 \\
    good                  & 0.60 & 0.39 & 10 \\
    \midrule
    overall               & 0.75 & 0.63 & 41 \\
    \bottomrule
  \end{tabular}
  \caption{Cohen's $\kappa$ per concept. \emph{Participant}: two extractions
  of the same participant texts, 157 cells over 32 participants, three
  unresolved cells excluded. \emph{Viewpoint}: re-labelling of the 47 target
  viewpoints against the corpus labels, 235 cells. \emph{Neutral}: how many of
  the 47 corpus labels for that concept are neutral. The two concepts the
  labellers agree on least are the two with the most neutral labels.}
  \label{tab:label-agreement}
\end{table}

Agreement is lowest on the same concept in both comparisons, \emph{good}, which is also the most abstract of the five. The two weakest concepts, \emph{good} and \emph{outdoor activity}, are the two whose corpus labels
contain the most neutral values. Counted over participant cells, where each of the 63 participants contributes the five concepts of their own target viewpoint, the two labellers differ on 62\% of the cells the corpus labels as neutral (50/81) against 12\% of polar cells (27/234). Disagreement is therefore concentrated on neutral stances, which is where participants were also least accurate.

\newpage
\section{Target Specificity}
\label{app:geometry}

For each participant, $g$ is the share of concepts on which the reconstruction matches the assigned viewpoint, minus the mean share over the four other viewpoints in the session. 
Because $g$ depends on how similar the five viewpoints are, we subtract its expected value under a reconstructor
with no knowledge of the assigned viewpoint, evaluated on the participant's own five viewpoints. Table~\ref{tab:geometry} reports the condition difference under three such reference reconstructors. Under uniform random answers the expected value is zero, so that row equals the raw difference.

\begin{table}[h]
  \centering
  \small
  \begin{tabular}{lccc}
    \toprule
    Reference & $n$ & $b$ (\textsc{Psv}$-$\textsc{Scalar}) & $p$ \\
    \midrule
    None (raw $g$)        & 63 & $+0.055$ [$-0.07$, 0.18] & .41 \\
    Participant's stance  & 59 & $+0.083$ [$-0.10$, 0.27] & .37 \\
    Session centroid      & 63 & $+0.043$ [$-0.09$, 0.18] & .52 \\
    Uniform random        & 63 & $+0.055$ [$-0.07$, 0.18] & .41 \\
    \bottomrule
  \end{tabular}
  \caption{Condition difference in target specificity, before and after adjusting for how far apart a session’s five viewpoints were. OLS with HC3 standard errors, one row per participant.}
  \label{tab:geometry}
\end{table}

More participants scored above zero on the target-specificity difference $g$ under \textsc{Psv} than under \textsc{Scalar}, 24/32 against 13/31. The mean difference in $g$ is small and its confidence interval includes zero, and adjusting for session geometry does not change this. The adjustment is needed because \textsc{Psv} sessions were more spread out, with a mean $L_1$ distance of 6.3 from the target to the other four viewpoints against 5.3 under \textsc{Scalar}, which makes the target easier to tell apart from its neighbours.

\end{document}